\documentclass[runningheads]{llncs}
\usepackage{graphicx}
\usepackage{footnote}
\usepackage{adjustbox}
\usepackage{mathtools}
\usepackage{tabularx}
\usepackage{longtable}
\usepackage{arydshln}
\usepackage{multirow}
\usepackage{booktabs}
\usepackage{xcolor}
\usepackage{comment}
\usepackage{graphicx}
\usepackage{hyperref}
\usepackage[position=bottom]{subfig}

\begin{document}
\title{EdNet: A Large-Scale Hierarchical Dataset in Education}
%
%
\author{Youngduck Choi\inst{1,2} \and
Youngnam Lee\inst{1} \and
Dongmin Shin\inst{1} \and
Junghyun Cho\inst{1} \and \\
Seoyon Park\inst{1} \and 
Seewoo Lee\inst{1,3} \and
Jineon Baek\inst{1,4} \and
Chan Bae\inst{1,3} \and \\
Byungsoo Kim\inst{1} \and
Jaewe Heo\inst{1}}
%
%
\institute{Riiid! AI Research \and Yale University \and UC Berkeley \and University of Michigan\\
\email{\{youngduck.choi,yn.lee,dm.shin,jh.cho,seoyon.park,seewoo.lee,\\jineon.baek,chan.bae,byungsoo.kim,jwheo\}@riiid.co}}
\maketitle              
\begin{abstract}
Advances in Artificial Intelligence in Education (AIEd) and the ever-growing scale of Interactive Educational Systems (IESs) have led to the rise of data-driven approaches for knowledge tracing and learning path recommendation.
Unfortunately, collecting student interaction data is challenging and costly.
As a result, there is no public large-scale benchmark dataset reflecting the wide variety of student behaviors observed in modern IESs.
Although several datasets, such as ASSISTments, Junyi Academy, Synthetic and STATICS are publicly available and widely used, they are not large enough to leverage the full potential of state-of-the-art data-driven models.
Furthermore, the recorded behavior is limited to question-solving activities.
To this end, we introduce \emph{EdNet}, a large-scale hierarchical dataset of diverse student activities collected by \emph{Santa}, a multi-platform self-study solution equipped with an artificial intelligence tutoring system.
\emph{EdNet} contains 131,417,236 interactions from 784,309 students collected over more than 2 years, making it the largest public IES dataset released to date.
Unlike existing datasets, \emph{EdNet} records a wide variety of student actions ranging from question-solving to lecture consumption to item purchasing.
Also, \emph{EdNet} has a hierarchical structure which divides the student actions  into 4 different levels of abstractions.
The features of \emph{EdNet} are domain-agnostic, allowing EdNet to be easily extended to different domains.
The dataset is publicly released for research purposes.
We plan to host challenges in multiple AIEd tasks with \emph{EdNet} to provide a common ground for the fair comparison between different state-of-the-art models and to encourage the development of practical and effective methods. 

\end{abstract}

\section{Introduction}
Knowledge tracing, the task of modelling a student's knowledge state through their learning activities over time, is a long-standing challenge of Artificial Intelligence in Education (AIEd).
Since understanding a student’s knowledge state is a crucial step for many problems of interest including learning path recommendation, score prediction and dropout prediction, knowledge tracing is considered one of the most fundamental problems of AIEd.

With advances in data science and the increasing availability of Interactive Educational Systems (IESs), data-driven models that learn the complex nature of student behaviors from interaction data have become a common recipe for knowledge tracing \cite{piech2015deep,zhang2017dynamic,huang2019ekt,pandey2019self,lee2019creating}.
However, the AIEd research community currently lacks a large-scale benchmark dataset which reflects the wide variety of student behaviors available through modern IESs.
Although several datasets, such as ASSISTments \cite{feng2009addressing,pardos2014affective}, Junyi Academy \cite{chang2015modeling}, Synthetic \cite{piech2015deep} and STATICS \cite{koedinger2010data}, are available to the public and widely used by AIEd researchers, they are not large enough to leverage the full potential of data-driven models.
Furthermore, the data they collect is limited to question-solving activities.

In this paper, we introduce \emph{EdNet}\footnote{\url{https://github.com/riiid/ednet}}, a large-scale hierarchical dataset consisting of student interaction logs collected over more than 2 years from \emph{Santa}\footnote{\url{https://santatoeic.com}}, a multi-platform, self-study solution equipped with artificial intelligence tutoring system that aids students in preparing for the TOEIC\textsuperscript{\textregistered} (Test of English for International Communication\textsuperscript{\textregistered}) test.
To the best of our knowledge, \emph{EdNet} is the largest dataset open to the public, containing 131,441,538 interactions from 784,309 students.
Aside from question-solving logs, \emph{EdNet} also contains diverse student behaviors including but not limited to self-study activities, choice elimination, and course payment.
\emph{EdNet} has a hierarchical structure where the possible student actions in \emph{Santa} are divided into 4 different levels of abstraction.
This allows the researcher to select the level appropriate for the AIEd task at hand, for example, knowledge tracing or learning path recommendation.
We release \emph{EdNet} to the public under the Creative Commons Attribution-NonCommercial 4.0 International license for research purposes only. 

\section{EdNet}
\begin{figure*}[t]
\centering
\includegraphics[width=1\textwidth]{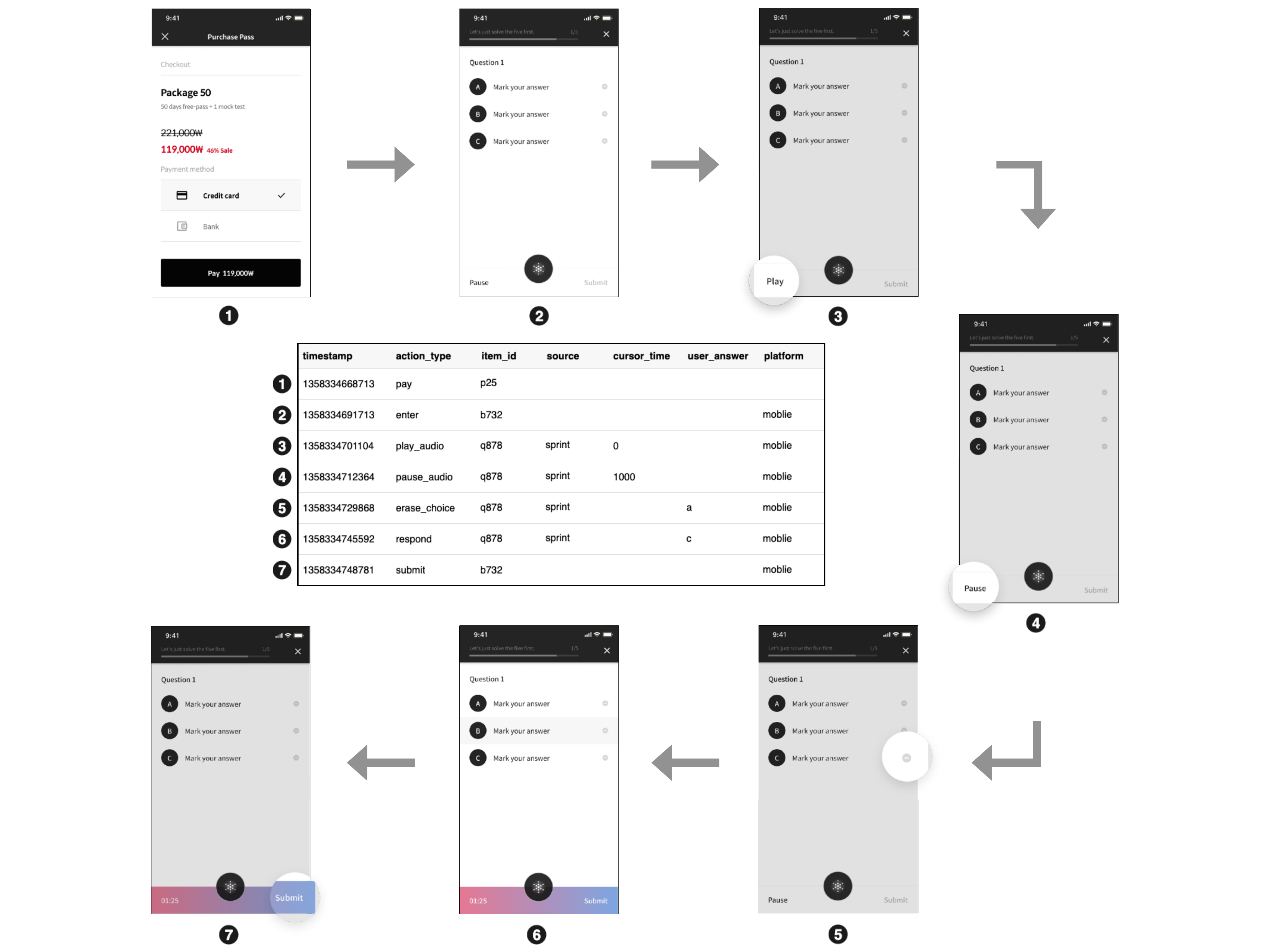}
\caption{A possible scenario of a student using \emph{Santa}.
After the student purchases a 50-day pass, they solve an LC question.
While they solve the question, all of their actions including audio playing and choice elimination are recorded.}
\label{fig:kt_example}
\end{figure*}

\emph{EdNet} is a dataset consisting of all student-system interactions collected over a period spanning two years by \emph{Santa}, a multi-platform AI tutoring service with approximately 780,000 students in South Korea.
\emph{Santa} is available through Android, iOS and the Web.
It aims to prepare students for the TOEIC (Test of English for International Communication\textsuperscript{\textregistered}) Listening and Reading Test.
Each student communicates their needs and actions through \emph{Santa}, to which the system responds by providing video lectures, assessing their response or giving expert commentary. 
\emph{Santa}'s UI and data-gathering process is described in Figure \ref{fig:kt_example}.
As shown in the figure, the \emph{EdNet} dataset contains various features of student actions such as the identity of the learning material consumed or the time spent by the student in solving a given problem.

\subsection{Properties of EdNet}
\subsubsection{Large-scale}
\emph{EdNet} is composed of a total of 131,441,538 interactions collected from 784,309 students of \emph{Santa} since 2017.
Each student has generated an average of 441.20 interactions while using \emph{Santa}.
Based on those interactions, \emph{EdNet} makes it possible for researchers to access large-scale real-world IES data.
Moreover, \emph{Santa} provides a total 13,169 problems and 1,021 lectures tagged with 293 types of skills, and each of them has been consumed 95,294,926 times and 601,805 times, respectively.
To the best of our knowledge, this is the largest dataset in education available to the public in terms of the total number of students, interactions, and interaction types.

\begin{table*}[t]
\caption{Comparison of Datasets in Education (ASSISTments \cite{feng2009addressing,pardos2014affective}, Synthetic-5 \cite{piech2015deep}, Junyi Academy \cite{chang2015modeling}, STATICS-2011 \cite{koedinger2010data} and EdNet).
Here \emph{logs} stands for interactions, and Synthetic-5, Junyi Academy, and Statics-2011 are renamed as Syn-5, Junyi, and Stat-2011.}
\centering
\begin{adjustbox}{width=\textwidth}
\begin{tabular}{lcccccccccc}
\toprule
            \multirow{ 2}{*}{}           & \multicolumn{3}{c}{ASSISTments}  & \multirow{ 2}{*}{Syn-5}  & \multirow{ 2}{*}{Junyi}  & \multirow{ 2}{*}{Stat-2011}  & \multicolumn{4}{c}{EdNet} \\ \cmidrule{2-4} \cmidrule{8-11}
                       & 2009 & 2012 & 2015 &  &  &  &  KT1 & KT2 & KT3 & KT4 \\
\midrule
\# of students          &        4,217       &        46,674        &   19,917  &            4,000        &    247,606    &    335 &   784,309   &   297,444      &    297,915     &   297,915     \\
\# of questions         &        26,688       &         179,999       &     100     &             50           &       722        &       1,362     &   13,169  &   13,169         &        13,169        &         13,169          \\
\# of tags       &         123      &          265      &               - &             5           &    41           &       27         & 188 &   188           &       293       &          293         \\
\# of lectures          &      0         &      0          &          0          &    0          &       0        &       0         &   0   &   0       &       1,021         &         1,021           \\
\# of logs      &       346,860        &        6,123,270        & 708,631 &              200,000 &      25,925,922        &      361,092      &  95,293,926  &   56,360,602    &  89,270,654     &   131,441,538   \\
\# of types of logs & 3 & 3 & 1 &  1 & 1&5 & 1 & 3 & 4 & 13 \\
Public available        &       Yes       &        Yes       &   Yes             &    Yes           &       Yes        &        No   &  Yes  &      Yes          &         Yes      &   Yes               \\
Contents available & No& No & No & No  &Yes &Yes &No &No &No &No \\
From real-world         &      Yes         &  Yes                    &   Yes              &    No            &         Yes     &       Yes         &   Yes     &  Yes      &       Yes         &  Yes                 \\
Collecting period       &     1y          &        1y        &        1y        &          -   &      2y 2m         &      4m          & 2y 7m &   1y 3m           &       1y 3m          &         1y 3m           \\
\bottomrule
\end{tabular}
\end{adjustbox}
\label{tab:comp}
\end{table*}

\subsubsection{Diversity}
\emph{EdNet} offers the most diverse set of interactions among all existing public IES datasets as can be seen in Table \ref{tab:comp}.
The set of behaviors directly related to learning is also richer in \emph{EdNet} than in other datasets, as \emph{EdNet} includes learning activities such as reading explanations and watching lectures which aren't provided in other datasets.
The richness of the data enables researchers to analyze students from various perspectives.
For example, purchasing logs may help analyze student's engagement with the learning process.

\subsubsection{Hierarchy}
\begin{figure*}[t]
\centering
\includegraphics[width=1\textwidth]{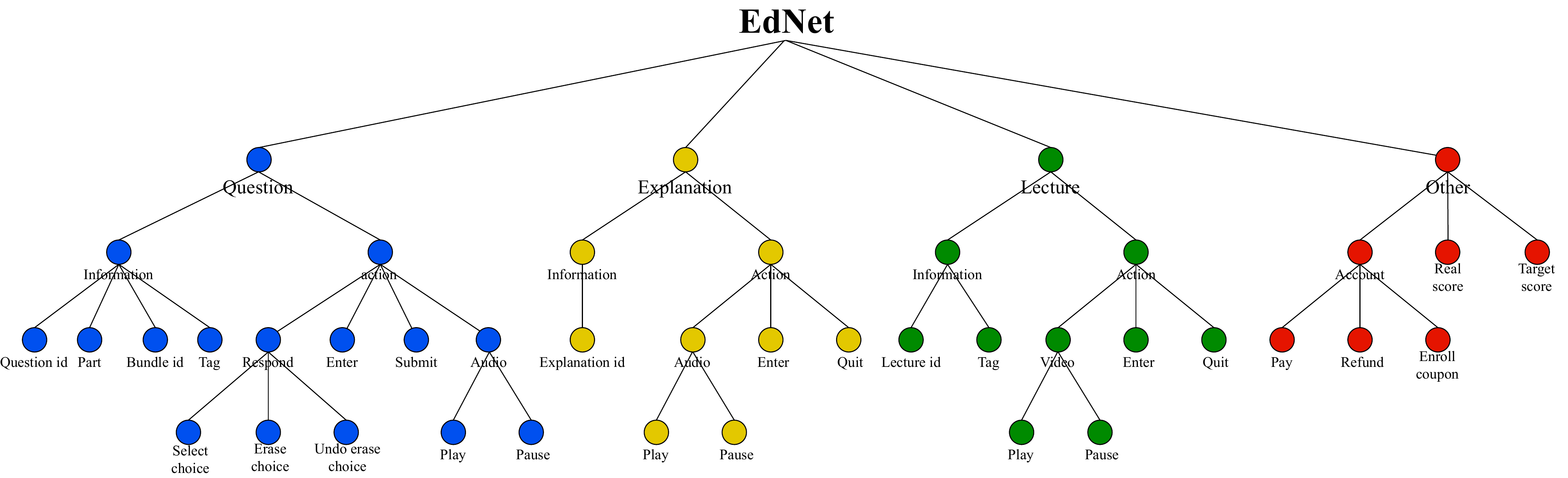}
\caption{Hierarchical structure of EdNet.}
\label{fig:EdNet_structure}
\end{figure*}

\emph{EdNet} is organized into a hierarchical structure where each level contains different types of data points as shown in Figure \ref{fig:EdNet_structure}. 
To provide the various types of data in a consistent and organized manner, \emph{EdNet} offers the data in four different datasets named KT1, KT2, KT3 and KT4.
As the postfix index of the datasets increases, the number of actions and types of actions involved also increase as shown in Table \ref{tab:comp}.
The details of each dataset is described in Section \ref{sec:hier}.

\subsubsection{Multi-platform}
In an age dominated by various devices spanning from personal computers to smartphones and AI speakers, IESs must offer access from multiple platforms in order to stay competitive. 
Accordingly, \emph{Santa} is a multi-platform system available on iOS, Android and Web and \emph{EdNet} contains data points gathered from both mobile and desktop users.
\emph{EdNet}'s platform-agnostic design allows the study of AIEd models suited for future multi-platform IESs, utilizing the data collected from different platforms in a consistent manner.

\subsection{Hierarchy of \emph{EdNet}}
\label{sec:hier}
The raw records obtained by \emph{Santa} accurately and thoroughly represent each student's learning process. 
However, the unprocessed details of raw records are hard to utilize directly for AIEd tasks and require pre-processing to extract meaningful information. 
In order to aid the process, we pre-process the collected records into four datasets of different levels of abstractions named KT1, KT2, KT3 and KT4.
The resolution of each dataset increases in the given order, starting from question-response interaction sequences used by most deep knowledge tracing models to the complete list of user actions gathered by \emph{Santa}.
The datasets were designed with particular tasks in mind so that one can use them readily for AIEd applications such as knowledge tracing, score prediction and dropout prediction. 
We describe each dataset of \emph{EdNet} as the following.

\subsubsection{EdNet-KT1}
In its simplest form, the learning session of a student can be described as a sequence of question-response pairs.
That is,
$$
(q_1, r_1), (q_2, r_2), \cdots, (q_t, r_t)
$$
where $q_i$ is the $i$'th question suggested by IES and $r_i$ is the student's response to $q_i$.
This is the format used by various deep-learning knowledge tracing models such as Deep Knowledge Tracing \cite{piech2015deep} and Self-Attentive Knowledge Tracing \cite{pandey2019self}.
EdNet-KT1 is the record of question-response pairs collected by \emph{Santa} from Apr. 18, 2017. 

We have organized \emph{EdNet} so that the questions come in bundles.
A bundle is a collection of questions sharing a common passage, picture or listening material. 
For example, questions with ids q2319, q2320 and q2321 may share the same reading passage. 
In that case, the questions are said to form a bundle and will be presented to the student at the same time along with the corresponding shared passage. 
When a bundle is given, a student must respond all the included questions in order to complete the bundle. 

\subsubsection{EdNet-KT2}
A major limitation of the question-response sequence format is that it is a very concise summary of student activities. 
For example, a student may alternate between one of two answer choices before deciding on one and submitting their final answer.
This potentially signals that they have narrowed the answer down to one of two choices but are unsure which one of the two are correct.
Modern IESs have the capability to log such details, but the question-response format cannot effectively represent such situation, limiting the analyses that can be performed with EdNet-KT1.
Since Aug. 27, 2018, \emph{Santa} has collected the full behavior of students to overcome this limitation. 
The resulting datasets, EdNet-KT2, EdNet-KT3 and EdNet-KT4, contain the compiled action sequences of each user. 
Each action represents a single unit of behavior.
They are made by students in the \emph{Santa} UI, and include such actions as watching a video lecture, choosing a response option, or reading a passage.
By recording a student's behavior as-is, the datasets represent each student's behavior more accurately and allow AIEd models to incorporate the finer details of a student's learning history.
EdNet-KT2, the simplest action-based dataset in \emph{EdNet}, consists of the actions related to question-solving activities.

\subsubsection{EdNet-KT3}
In \emph{Santa}, a student may participate in various learning activities other than solving questions.
These include reading through expert commentaries or watching lectures provided by the system.
EdNet-KT3 incorporates information on these learning activities.
This information can be utilized to infer the impact of learning activities on each student's knowledge state. 
For example, one could analyze the time each student spends studying certain expert commentaries and observing its effect on different learning behaviors and performance.

\begin{table*}[t]
\caption{Example student data in EdNet-KT1.
The student solved 3 bundles of total 5 questions.}
\centering
\begin{tabular}{cccccc}
\toprule
timestamp     & solving\_id& question\_id  & user\_answer & elapsed\_time \\
\midrule
1548996377530 & 48      & q2844     & d         & 47000         \\
1548996378149 & 48      & q2845     & d         & 47000         \\
1548996378665 & 48      & q2846     & d         & 47000         \\
1548996671661 & 49      & q4353     & c         & 67000         \\
1548996787866 & 50      & q3944     & a         & 54000         \\
\bottomrule
\end{tabular}
\label{tab:kt1}
\end{table*}

\begin{table*}[t]
\caption{Example student data in KT2.
The student solved 4 questions on mobile, 3 of which were contained in the bundle b5180.}
\centering
\begin{tabular}{cccccc}
\toprule
timestamp     & action\_type & item\_id & source    & user\_answer & platform \\
\midrule
1540872586641 & enter        & b4957    & diagnosis &              & mobile   \\
1540872610841 & respond      & q6425    & diagnosis & c            & mobile   \\
1540872624073 & leave        & b4957    & diagnosis &              & mobile   \\
1540872626743 & enter        & b5180    & sprint    &              & mobile   \\
1540872675318 & respond      & q6815    & sprint    & c            & mobile   \\
1540872692681 & respond      & q6816    & sprint    & a            & mobile   \\
1540872694917 & respond      & q6814    & sprint    & a            & mobile   \\
1540872700262 & leave        & b5180    & sprint    &              & mobile   \\
\bottomrule
\end{tabular}
\label{tab:kt2}
\end{table*}

\begin{table*}[t]
\caption{Example student data in KT3.
After the student solved question q790, they read the explanation associated with q790 and watched lecture l540.}
\centering
\begin{tabular}{cccccc}
\toprule
timestamp     & action\_type & item\_id & source          & user\_answer & platform \\
\midrule
1573364188664 & enter        & b790     &  sprint           &              & mobile   \\
1573364209673 & respond      & q790     &  sprint           & d            & mobile   \\
1573364209710 & leave        & b790     &   sprint          &              & mobile   \\
1573364209745 & enter        & e790     &   sprint          &              & mobile   \\
1573364218306 & leave        & e790     &   sprint          &              & mobile   \\
1573364391205 & enter        & l540     &  adaptive\_offer  &              & mobile   \\
1573364686796 & leave        & l540     &  adaptive\_offer  &              & mobile   \\
\bottomrule
\end{tabular}
\label{tab:kt3}
\end{table*}

\begin{table*}[t]
\caption{Example student data in EdNet-KT4.
After the student bought an item, they solved the LC question q878.
The timestamps at which they played and paused audio were recorded.
They also eliminated `a' and chose `c' as an answer.}
\centering
\begin{tabular}{ccccccc}
\toprule
timestamp & action\_type   & item\_id   & source& cursor\_time & user\_answer & platform  \\
\midrule
1358114668713   & pay               & p25   &           &       &   & mobile    \\
1358114691713   & enter             & b878  & sprint    &       &   & mobile    \\
1358114701104   & text\_enter       & q878  & sprint    &       &   & mobile    \\
1358114712364   & audio\_play       & q878  & sprint    & 0     &   & mobile    \\
1358114729868   & audio\_pause      & q878  & sprint    & 10000 &   & mobile    \\
1358114745592   & eliminate\_choice & q878  & sprint    &       & a & mobile    \\
1358114748023   & select            & q878  & sprint    &       & c & mobile    \\
1358114748781   & leave             & b878  & sprint    &       &   & mobile    \\
\bottomrule
\end{tabular}
\label{tab:kt4}
\end{table*}

\subsubsection{EdNet-KT4}
In EdNet-KT4, the complete list of actions collected by \emph{Santa} is provided. 
The purpose of EdNet-KT4 is to provide the very fine details of student activity as recorded by \emph{Santa}.
This allows access to features and tasks specific to a particular design.
For example, one may analyze the impact that purchasing a paid course has on the studying behavior of a student.

\section{Related Datasets}
\subsection{Synthetic}
Synthetic is a public dataset\footnote{https://github.com/chrispiech/DeepKnowledgeTracing/tree/master/data/synthetic} made by the authors of Deep Knowledge Tracing \cite{piech2015deep}. 
Based on Item Response Theory (IRT), a total of 4K virtual students answering a fixed set of 50 questions drawn from $k \in \{1, \dots, 5\}$ concepts are generated.
More precisely, each student has a "skill" for each concept represented by a single number $\alpha$, and each exercise has a difficulty represented by a number $\beta$. 
The probability that the student answer a question correctly is modelled by the 3PL model of IRT given as $p(\mathrm{correct} | \alpha, \beta) = c + \frac{1-c}{1 + e^{\beta-\alpha}}$, where $c$ is the probability of student guessing randomly (which is $0.25$, as we assume that there are 4 choices for each question).
Two datasets with $k=2, 5$ are publicly available. 

\subsection{ASSISTments}
ASSISTments datasets \cite{feng2009addressing,pardos2014affective} are collected from the ASSISTment system, an online tutoring system which provides instructional assistance while assessing students at the same time.
Each ASSISTment consists of an original question from Massachusetts Comprehensive Assessment System (MCAS) 8th math test items, and a list of scaffolding questions created by domain experts.
For each ASSISTment, a student first attempts an original question.
If the student fails to answer the original question, scaffolding questions unfolds to guide the student towards the sub-steps required to solve the original question correctly.
While solving the scaffolding questions, the student can give answers to the questions or ask for help-seeking behavior.
Depending on the student actions, the system gives suitable feedbacks such as hints, buggy messages or answers to the scaffolding questions.

\subsection{STATICS2011}
The STATICS2011 dataset contains 335 engineering student's question-solving logs from a one-semester statics course via online educational system developed by Carnegie Mellon University. 
The dataset can be found in PSLC datashop \cite{koedinger2010data}, which is not public but can be obtained by request\footnote{https://pslcdatashop.web.cmu.edu/DatasetInfo?datasetId=507}.
The dataset is used by several works on knowledge tracing, such as DKVMN \cite{zhang2017dynamic}, SAKT \cite{pandey2019self}, and SKVMN \cite{abdelrahman2019knowledge}. 

\subsection{Junyi Academy}
Junyi Academy\footnote{http://www.junyiacademy.org/} is an e-learning platform in Taiwan that provides 722 mathematics questions. 
The Junyi Academy dataset was first introduced in \cite{chang2015modeling}, which is also available online in PSLC datashop\footnote{https://pslcdatashop.web.cmu.edu/DatasetInfo?datasetId=1198}. 
For each question, there are two kinds of information that can be considered as tags - \texttt{topic} and \texttt{area}. 
There are 40 different kinds of \texttt{topics} (except \texttt{nan} denoting the absence of topic) that are assigned to each question, such as \texttt{absolute-value}, \texttt{circle-properties}, and \texttt{fractions}. \texttt{area} can be considered as a more general concept of \texttt{topic}, where each \texttt{area} contains several \texttt{topics}. 
There are a total of 7 areas (except \texttt{nan}) including \texttt{arithmetic}, \texttt{logics}, and \texttt{algebra}.
It also provides question information data, such as knowledge map, prerequisites, expert-annotated time limit and similarity between two exercises, separately.  
This dataset has been is widely used for various educational tasks including knowledge tracing \cite{abdelrahman2019knowledge} and learning path recommendation \cite{liu2019exploiting}. 

\subsection{PSLC Datashop}
PSLC (Pittsburgh Science of Learning Center) datashop\footnote{https://pslcdatashop.web.cmu.edu/} is an open data repository maintained by Carnegie Mellon University \cite{koedinger2010data}. 
Various learning interaction datasets including STATICS2011, Junyi Academy datasets and also Elementary Chinese Course and Intelligent Writing Tutor are available.

\section{Applications of EdNet}
In this section, we show three applications of \emph{EdNet}: knowledge tracing, study session dropout prediction in mobile learning environments, and pre-training tasks for label-scarce problems in AIEd.
We also discuss the possibility of developing student simulators with \emph{EdNet} with an eye towards applications in Reinforcement Learning(RL)-based methods for learning path recommendation.

\subsection{Knowledge Tracing}
Knowledge tracing is the problem of modeling a student’s knowledge state from their actions in learning activities.
The recently published \textbf{S}eparated Self-\textbf{A}ttent\textbf{I}ve \textbf{N}eural Knowledge \textbf{T}racing (SAINT) is a Transformer-based \cite{vaswani2017attention} knowledge tracing model trained on EdNet-KT1 \cite{choi2020appropriate}. 
Like many other existing knowledge tracing models, SAINT tackles the classical problem of predicting student response correctness to each exercise, given the history of responses to previous exercises.
Compared to previous knowledge tracing models, SAINT achieved state-of-the-art performance with an dramatic improvement in Area Under receiver operating characteristic Curve (AUROC).
The large scale of EdNet-KT1 data allows the model to capture complex relationships between student interactions through deep attention layers.

\subsection{Study Session Dropout Prediction in Mobile Learning Environments}
\cite{lee2020deep} is the first work investigating the study session dropout prediction problem in mobile learning environments.
This work defined a study session as a sequence of learning activities where the time interval between adjacent activities is less than 1 hour.
Accordingly, study sessions are shorter in time, and occurs more frequently than the course dropout.
Based on the definition, \cite{lee2020deep} proposed a Transformer-based \textbf{D}eep \textbf{A}ttentive \textbf{S}tudy Session Dropout Prediction model (DAS).
Empirical evaluations on EdNet-KT4 showed that DAS outperformed existing models.

\subsection{Pre-training Tasks for Label-scarce Educational Problems}
Label-scarcity is a challenging issue in many subfields of AI, and AIEd is no exception.
For instance, standardised exam scores cannot be obtained within IESs.
The researcher must take action outside the IES to obtain them.
Furthermore, some features available through IESs, including course dropout and review correctness, tend to be few in number since they occur sporadically in practice.
To this end, \cite{choi2020assessment} proposed Assessment Modeling, a pre-training method for general IESs.
\cite{choi2020assessment} defined the notion of an assessment, a specific type of educational interactive features available in IESs, and proposed using assessment prediction as a pre-training task for label-scare AIEd tasks.
Using EdNet-KT4 as the training dataset, Assessment Modeling showed state-of-the-art results in both exam score and review correctness prediction, outperforming pre-training methods developed in the natural language processing community that learn representations of the contents of learning items. 

\begin{figure*}[t]
\centering
\includegraphics[width=1\textwidth]{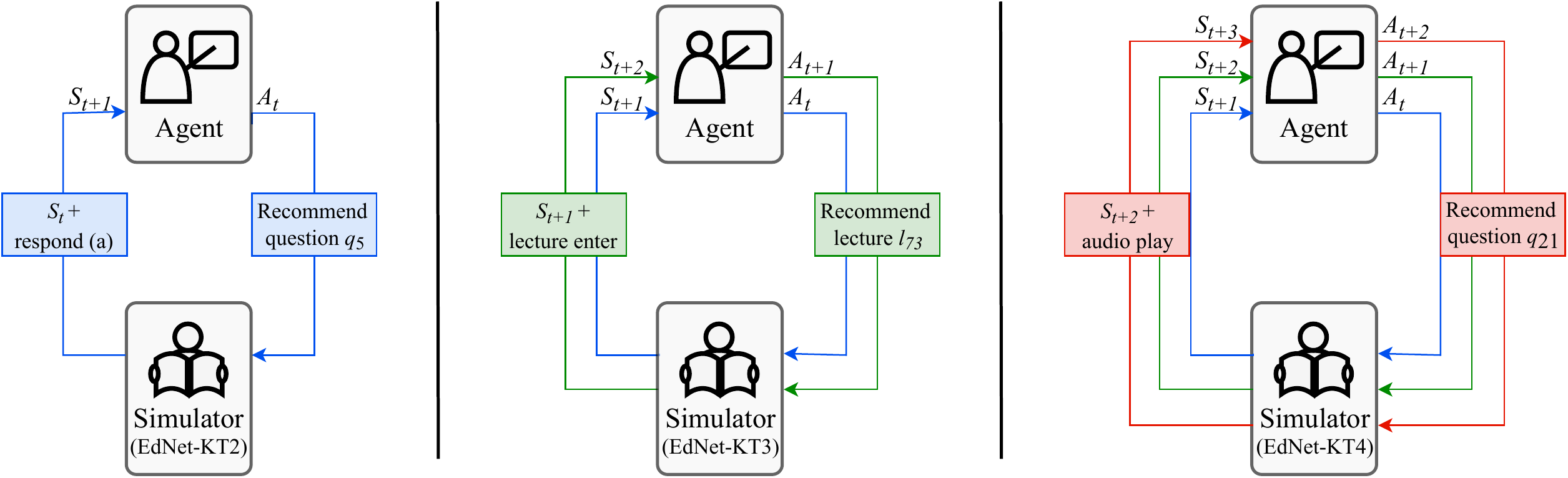}
\caption{Simulating students at various levels of detail. Here $A_{t}$ and $S_{t}$ refer to the action and the state at $t$-th step, respectively. The number of actions that agents can offer and responses that the simulator can generate increase as the level of the dataset does.}
\label{fig:simulator}
\end{figure*}

\subsection{RL-based Learning Path Recommendation}
Even if a student's knowledge state is fully understood, there still remains the task of providing an individualized, optimal learning path.
In this regard, educational content recommendation has been studied extensively \cite{xu2016personalized,pardos2019designing} with Reinforcement Learning (RL) emerging as a prominent method \cite{liu2019exploiting,huang2019exploring,zhou2019hierarchical,shawky2018reinforcement,reddy2017accelerating,mandel2017better}.
In the context of RL, a policy (e.g. tutoring strategy) is trained to maximize the reward function that evaluates the overall educational effect of the agent (tutor) over time.
Most RL algorithms require repeated evaluations of the agent for training.
As evaluating a tutoring strategy with real student is extremely costly, most methods use simulated students for training \cite{liu2019exploiting,huang2019exploring}. 

\emph{EdNet} offers multiple levels of features from which simulators of varying levels of fidelity can be developed (see Figure \ref{fig:simulator}). 
For instance, one may build a simulator that generates a virtual student's response to newly suggested questions by training a model that predicts their response from their question-solving history with EdNet-KT1.
With EdNet-KT4, more detailed actions such as lecture-watching, product purchase or answer choice elimination can be simulated in the same way. 
Each option trades off simplicity with fidelity. 
By allowing a range of options for this trade-off, \emph{EdNet} gives the researcher the opportunity to find the most appropriate simulator that suits the particular goal of the agent being trained.

\section{Conclusion}
In this paper, we introduced \emph{EdNet}, a large-scale dataset in education gathered by the multi-platform service \emph{Santa}.
\emph{EdNet} contains a very high-resolution record of each user's activities and, to date, it is much larger than any other public dataset in education. 
The hierarchical structure of \emph{EdNet} allows researchers to approach diverse tasks in AIEd from various levels of abstraction. 
We believe that \emph{EdNet} will provide fertile soil for further developments in AIEd, and we will continue to keep the dataset up-to-date.

\section{Acknowledgement}
The authors would like to thank all the members of \emph{Riiid!} for leading the \emph{Santa} service successfully.
\emph{EdNet} could not have been compiled without their efforts.

\bibliographystyle{splncs04}
\bibliography{references.bib}

\end{document}